\newcommand{\bee}{\begin{equation}}
\newcommand{\ene}{\end{equation}}
\newcommand{\beea}{\begin{eqnarray}}
\newcommand{\enea}{\end{eqnarray}}

\documentclass[aps,preprint,pre,showpacs,superscriptaddress]{revtex4}
\usepackage{amsmath} 
\usepackage{graphicx}
\usepackage{dcolumn}
\usepackage{bm}
\begin{document}
\title{The Energy Dispersion of Matter in Schr\"{o}dinger-Poisson Model}
\author{M. Akbari-Moghanjoughi}
\affiliation{Faculty of Sciences, Department of Physics, Azarbaijan Shahid Madani University, 51745-406 Tabriz, Iran}

\begin{abstract}
The energy dispersion of collective excitations of free electron gas as well as self-gravitating ensemble of uncharged particles is derived using a unified 1D Schr\"{o}dinger-Poisson model. The energy dispersion of self-gravitating system is shown to lead to unique features which are absent in the case of electrostatic excitations. Current mathematical model of collective particle excitations is shown to gives rise to a novel description of the paradoxical wave-particle duality and many intriguing new collective features due to the scale-dependence effective forces in collective interactions, absent in the single particle description of physical systems. It is shown that the excitations in a self-gravitating system lead to a fundamentally different features of collective interaction under gravitational potential than that of electrostatic ones. Particularly, it is found that the total energy of self-gravitating systems can be negative and the effective mass of excitations vary significantly in the whole spectrum of wavelength. The later may be considered as meaningful absence of self-consistent theory of quantum gravity. The significance of the peculiar aspects of the two fundamentally different excitation types is discussed based on their relevance to modern theories.
\end{abstract}

\date{\today}

\maketitle

Collective interactions are of fundamental importance in the study of statistical systems like plasmas, metals, semiconductors and Bose-Einstein condensates \cite{chen, ash, landau}. Due to a large degree of freedom in realistic models the number of appropriate analytical tools for investigation of the important collective excitations is strongly limited. The complexity of many body systems becomes even manifold as the quantum density limit is approached \cite{bohm,pines}. However, many innovative tools have been developed in order to cope with this complexity usually by numerical means. These tools range from the many-body Hartree-Fock approximation, Vlasov-Poisson and Wigner-Poisson kinetic theories \cite{manfredi} to the Kohn-Sham density functional variations \cite{bonitz}, to name a few. On the other hand, quantum hydrodynamic (QHD) models \cite{haasbook,moldabekov}, although not as accurate as kinetic ones, are common tools which provide some analytical results. Recently, QHD has been used to reveal some interesting collective aspect of quantum plasmas \cite{shukla}. The Schr\"{o}dinger-Poisson model has recently been used to numerically study full nonlinear collective effects such as propagation and interactions of dark solitons \cite{se} and plasmon harmonic generation \cite{hurst} in quantum plasmas accounting for the self-consistent mean-field Hartree potential caused by the collective charge interactions.

On the other hand, a simple analytical extension of the linear single-particle Schr\"{o}dinger model accounting for basic collective interactions seems to be the next straightforward move towards a meaningful understanding of collective quantum phenomena. Such a theory may also provide correct answers to some fundamental questions like the quantum measurements and wave-particle duality \cite{zm,leg} targeting the very basic concepts of the quantum theory itself. Moreover, an appropriate full-featured quantum theory may lead to more appropriate theory of quantum cosmology which does not bear the shortcomings of the standard model \cite{matt,neil}. Recent N-body numerical simulation \cite{man} of the self-gravitating system with negative mass particles based on the Vlasov-Poisson model reveal different possible structure formation and expansionn scenarios which are scale dependent. The negative mass of particles of hypothetical dark matter of unknown origin which constitutes over $85$ percent of the Universe is usually attributed to antimatter in the Dirac-Milne symmetric matter-antimatter model \cite{levy}. However, current unified model for one-dimensional collective quantum excitations valid for both electrostatic as well as self-gravitating particles, nicely, describes the negative energy and effective mass needed for the expansion of the Universe to work. It is the coupling between the single and collective oscillations in current model however which quite analogous to starling murmuration leads to some interesting new physical phenomena.

Here, we would like to study both electrostatic and self-gravitating collective excitations in system of one-component species in a single one-dimensional mathematical model. Therefore, we consider the following effective Schr\"{o}dinger-Poisson equations \cite{manfrediprb}
\begin{equation}\label{sp}
- i\hbar \frac{{\partial {\cal N} }}{{\partial t}} = \frac{{{\hbar ^2}}}{{2m}}\frac{{\partial ^2 {{\cal N}}}}{{\partial {x^2}}} \pm \Phi {\cal N}  + \mu {\cal N} ,\hspace{3mm}\frac{{\partial ^2 {\Phi}}}{{\partial {x^2}}} = 4\pi \gamma{n_0}{{\cal N}^2},
\end{equation}
in which ${\cal N}={\cal N}/\sqrt{n_0}$ is the normalized density potential with $n_0$ being the equilibrium number density of species and $\Phi=\{e\phi_e,m\phi_g\}$ are the normalized electrostatic and gravitational potentials, respectively, with plus/minus sign corresponding to the electrostatic/self-gravitating system and $\mu$ being the chemical potential of the electrostatic system. Also, the constant $\gamma$ in Poisson's equation values $\gamma=\{K_e e^2,Gm^2\}$ for the electrostatic/self-gravitating systems with $K_e$ and $G$ being the electrostatic and gravitational constants, respectively. Since, we are looking for the stationary solutions we consider the separation of the variables as ${\cal N}(x,t)={\Psi(x)}{\cal E}(t)$, leading to the following system
\begin{subequations}\label{sps}
\begin{align}
&\frac{{{\hbar ^2}}}{{2m}}\frac{{\partial ^2 {\Psi}}}{{\partial {x^2}}} \pm \Phi \Psi  + \mu \Psi  = -\epsilon\Psi,\\
&i\hbar \frac{{\partial {\cal E}}}{{\partial t}} =  - \epsilon{\cal E},\\
&\frac{{\partial ^2 {\Phi}}}{{\partial {x^2}}} = 4\pi \gamma {n_0}{\Psi ^2}.
\end{align}
\end{subequations}
Now, using the wavefunction components, ${\cal N}(x,t)=\Psi(x)\exp(-i\epsilon t/\hbar)$ and $\Phi\equiv\Phi(x)$, we arrive at the following linearized coupled pseudoforce model
\begin{subequations}\label{pf}
\begin{align}
&\frac{{d^2{\Psi(x)}}}{{d{x^2}}} \pm \Phi(x) = - 2 E \Psi(x),\\
&\frac{{d^2{\Phi(x)}}}{{d{x^2}}} - {\Psi}(x) = 0,
\end{align}
\end{subequations}
in which we have used the definition for the normalized energy eigenvalues, $E=(\epsilon-\mu_0)/2E_0$ where $E_0=\hbar\omega_0$ with $\omega_0=\sqrt{4\pi \gamma n_0/m}$ being the plasmon/gravity quantum energy and $x=x/\lambda_0$ where $\lambda_0=2\pi/k_0$ with $k_0=\sqrt{2m E_0}/\hbar$ being the fundamental characteristic electrostatic/gravity length characterizing the collective interaction range. Note that for self-gravitating system with the equilibrium value of the chemical potential $\mu_0=0$ the current model is valid for any arbitrary value of specie mass, $m$, be it an atom in a neutral gas or a galaxy in the cosmos. As of the electrostatic excitations here we only consider the case for plasmon (the free electron collective excitations) of an arbitrarily degenerate free electron gas with $m=m_e$ being the free electron mass.

The general solution to the system (\ref{pf}) and the grand wavefunction characterizing the electrostatic and self-gravitating system of species can be written as follows
\begin{equation}\label{wf}
{\cal G}(x,t) =\left[ {\begin{array}{*{20}{c}}
{\Phi (x)}\\
{\Psi (x)}
\end{array}} \right] e^{-i\omega t} = \frac{e^{-i\omega t}}{{2\alpha }}\left[ {\begin{array}{*{20}{c}}
{{\Psi _0} + k_2^2{\Phi _0}}&{ - \left( {{\Psi _0} + k_1^2{\Phi _0}} \right)}\\
{ - \left( {{\Phi _0} \pm k_1^2{\Psi _0}} \right)}&{{\Phi _0} \pm k_2^2{\Psi _0}}
\end{array}} \right]\left( {\begin{array}{*{20}{c}}
{{{\rm{e}}^{\pm i{k_1}x}}}\\
{{{\rm{e}}^{i{k_2}x}}}
\end{array}} \right),
\end{equation}
\begin{equation}\label{eks}
{k_1} = \sqrt {E - \alpha },\hspace{3mm}{k_2} = \sqrt {E + \alpha },\hspace{3mm}\alpha  = \sqrt {{E^2} \pm i^2},
\end{equation}
where $\Phi_0$ and $\Psi_0$ are arbitrary constants characterizing the initial state of the system at origin. Note that the plus/minus signs correspond to the electrostatic/self-gravitating case. Note also that the relation $k_1 k_2=\sqrt{\pm1}$ holds in general. The solution (\ref{wf}) represents the right-going wave solution and the left-going wavefunction is obtained by replacing $i\to -i$.

It is evident from (\ref{wf}) that the grand wavefunction unlike the ordinary wavefunction of the Schr\"{o}dinger equation is characterized by two wavenumbers $k_1$ and $k_2$ due to two different length-scales in the considered particle systems. The total kinetic energy of the system being $E=k_1^2/2+k_2^2/2$ in which the energy and wavenumbers are normalized to $E_0$ and $k_0$, respectively. It is seen that the total energy is the composition of energies due to the single-particle and collective oscillations, leading to the energy dispersion relation $E(k)=(k^4\pm 1)/2k^2$. For instance, in the case of plasmon excitations, using the relation $k_1k_2=1$, we have $E=1/2k^2+ k^2/2=(k^4+1)/2k^2$ with the first term characterizing the single particle parabolic dispersion and the second term is new component to the plasmon dispersion due to Langmuir excitation of electron gas. The comparison of the contributing terms to the total energy reveals the fact that for $k>1$, that is, for length-scales less than $\lambda_0$ the contribution of single particle oscillation to the total energy is larger. This is when the single particle character of oscillations are probed. On the other hand, for $k<1$ the contribution from collective excitations to the total energy of plasmon excitations are larger compared to that of the single particle oscillations. Note that the identity $k_1k_2=1$ is a generalized complementarity relation, since, the wavenumbers $k_1$ and $k_2$ respectively characterize wave and particle aspects of plasmon excitations. It is concluded that for the length-scales much larger/smaller than $\lambda_0$ the wave/particle aspects of the plasmon excitation is revealed. However, for the length-scales comparable to $\lambda_0$, i.e., $E\simeq 1$ where $k_1\simeq k_2$ both wave and particle aspects merge. The later effect which is clearly contrary to the standard complementarity principle, avoiding simultaneous wave-particle manifestations, becomes possible when the quantum beating phenomenon takes place as the collective excitation wavelength approaches to that of the single particle oscillation.

On the other hand, for self-gravitating system of particles the energy dispersion relation becomes, $E=(k^4-1)/2k^2$. The identity $k_1 k_2=i$ dictates that either one of the wavenumbers corresponding to single particle or collective excitations is always imaginary. Figure 1(a) shows the plot of the energy dispersion relation for free particle (thin parabola) and the self-gravitating system which consists of real (solid) and imaginary (dashed) curves. It is remarked that, contrary to the case of electrostatic excitations, the total energy of a self-gravitating system can be negative. Also, for length-scales smaller/larger compared to the characteristic length $\lambda_0$ the wavenumber corresponding to the wave/particle branch becomes imaginary. This strongly suggests that the collective excitations in a self-gravitating system of particles always give rise to negative energies and imaginary wavenumber values, as also evident from the energy dispersion relation. This is to say that, contrary to the case of plasmon excitations, for the self-gravitating system collective entity always absorbs energy from the system of particles. The later difference between electrostatic and gravitational particle systems is the most fundamental aspect of the gravity leading to peculiar distinction of this force from others. Moreover, Fig. 1(b) schematically compares the dispersions due to electrostatic (thin curve) and self-gravitating (thick curves) systems. It is remarked that, for $k\gg 1$ the energy dispersion due to both excitations coincide with that of a free particle. The later occurs for length scales much smaller than the characteristic length, $\lambda_0$. The plasmon excitations have a minimum excitation energy value of $E_{min}=2E_0$ around which the quantum beating effect takes place when $k_1\simeq k_2$.

Figure 1(c) depicts the phase (dashed) and group (solid thin curve) speeds together with the fractional effective mass (thick curve) of plasmon excitations. In the free electron limit $k\gg 1$ the linear dependence of wave speeds with the characteristic free-particle relation $v_p=v_g/2$ are evident. Also, the plasmon mass approaches that of a single free electron in the limit of very small length-scales  compared to characteristic length $\lambda_0$, containing only a single electron. However, at very large-scales, the plasmon phase speed $v_p=E/p=(k^4\pm 1)/2\hbar k^3$ with $p=\hbar k$ being the de Broglie momentum and group speed $v_g=dE/dp=2(k/\hbar-v_p)$ substantially deviate from that of the single free electron. The phase speed of plasmon excitations have a minimum value of $k=3^{1/4}$. Note that the identity $\hbar(v_g+2v_p)=2k$ holds for both non-relativistic electrostatic and gravitational excitations. Moreover, the effective plasmon mass $m^*=m_0/(1\pm 3/k^4)$ with $m_0$ being the single free-electron mass, obtained using $1/m^*=d^2E/dp^2$, decreases as the length-scale increases from a single electron scale towards the full-scale at $k\to 0$, where the effective mass drops to zero. The concept of electron effective mass is more common in electron transport phenomenon in semiconductors. Figure 1(d) shows the wave speeds as well as the effective mass of self-gravitating system of uncharged particles. It is seen that, while the free electron limit of self-gravitating particle system shows similar effects, the large-scale character of wave speeds and the fractional effective mass prove to be fundamentally different from those of electrostatic excitations. In the case of gravity system the group speed has a minimum value at $k=3^{1/4}$ where the effective mass surprisingly diverges. This is the point of zero acceleration according to the generalized second law of Newton $F=m^* a$ where $m^*$ is the effective inertial mass, related to the single-particle mass via $m^*=m_0/(1- 3/k^4)$, which clearly depends on the number density of particle ensemble. It is also observed that below the diverging mass point at $k=3^{1/4}$ the effective mass becomes negative. Some simple analogy may be found for the negative mass electron in the semiconductors which is a manifestation of the free electron interaction with the lattice potential of a metallic or semiconductor crystal \cite{ash}. The concept of the negative mass and energy, on the other hand, may be one of the key ingredients of the modern cosmology. These concepts have been considered essential to explain the origin of dark matter and energy which are assumed to effectively rule the expansion of the large scale Universe \cite{man}.

The close inspection of the grand wavefunctions in (\ref{wf}) reveals some interesting fundamental properties of the collective electrostatic and gravitational excitations. For instance it is found that standing waves can be easily constructed from the generalized plane wave solutions of the plasmon excitations. The first quantization of free electron plasmon excitations with arbitrary degree of degeneracy has been demonstrated in Ref. \cite{akb}. In the case of self-gravitating particle system, however, this is impossible due to the fact that one of the wavenumbers is imaginary. This may be a reason for the meaningful absence of the long-south quantum theory of gravity. The wavefunction profiles are shown for collective plasmon and self-gravitating systems for different excitation energy values in Fig. 2. Figure 2(a) depicts the spacial variation plasmon grand wavefunction profiles for energy value close to the its minimum at quantum beating condition. It is remarked that probability and electrostatic potential wavefucntions are everywhere $180$ degrees out of phase. The increased energy value in Fig. 2(b) reveals the two tone character of the plasmon excitations which are due respectively to the single- and many-electron oscillations. Moreover, Fig. 2(c) shows the grand wavefunction profiles for self-gravitating system of uncharged particles. Both the probability and gravitational potential wavefunctions show growing oscillatory behavior in space for the positive energy value. The growing probability is an indication of the matter flow which is an interesting feature gravitating system. Calculation of the probability current $J(x)=(i\hbar/2m)[\Psi(x)\partial\Psi^*(x)/\partial x-\Psi^*(x)\partial\Psi(x)/\partial x]$ for the even and odd solutions of plasmon wavefunction results in zero probability current, while, the probability current for odd solution of the self-gravitating system is nonzero as follows
\begin{equation}\label{pc}
{J_{odd}}(x) = \frac{{{{\left( {1 + {k^2}} \right)}^2}}}{{{k^3}}}\left[ {{k^2}\cos (kx)\sinh \left( {\frac{x}{k}} \right) - \cosh \left( {\frac{x}{k}} \right)\sin (kx)} \right],
\end{equation}
which further increases spatially. In the Madelung fluid representation $J=\rho V$ where $\rho$ and $V$ represent the fluid mass density and velocity indicate a net flux of mass density which may be interpreted as the cosmological expansion in current model. Note that in the Bohmian deterministic interpretation the statistical ensemble of particles with a given number-density move with a speed which is proportional to the probability current \cite{bohmian}. Such interpretation is clearly a theoretical confirmation of the Hubble–Lemaître law \cite{hubble}. Therefore, current model may be considered as a theoretical model which explains the expansion of gravitating particle system without need for a hypothetical BIGBANG. Finally, Fig. 2(d) shows the wavefunction profiles for negative value of energy. It is remarked that for very large-scale excitations, where the length-scale exceeds that of the characteristic length, $\lambda_0$, both wavefunction profiles are still growing but non-oscillatory.

To conclude, using the Schr\"{o}dinger-Poisson model we investigated the grand wavefunction of electrostatic and self-gravitating particles in the framework of a unified coupled pseudoforce model. The presented model includes both contributions from single particle as well as collective effects in the grand wavefunction of the system of particles. It was revealed that collective effects and their fundamental contribution to the energy dispersion relation of both charged and uncharged systems leads to many new features in the dynamical properties of such systems for length scales comparable or larger than the characteristic length which depends on the mass and number-density of constituent species. Our study also showed that there are fundamental differences between the characteristics of solutions for electrostatic and self-gravitating systems due to the profound difference in the nature of involving interacting forces which prohibits the first quantization of a self-gravitating statistical ensemble similar to the case for charged particle systems like plasmas. Moreover, the energy dispersion relation of gravitational system of particles give rich and interesting insight into the nature of collective gravitational interactions which are highly significant in development of the more advanced theory of cosmology.

\end{document}